\documentclass{article}
\usepackage{spconf,amsmath,graphicx,hyperref,amssymb}

\title{Reference-Free Singing Pitch Correction via Music-Constrained Sequence Editing}
\name{\shortstack{Biao Dong$^{1}$, Jiajun Li$^{1}$, Binzhen Zhu$^{1}$, Mingwei Yi$^{1}$\\
Tong Liu$^{2}$, Yuanhao Zhang$^{2}$, Jiqing Han$^{1}$, Yongjun He$^{1}$}}
\address{$^{1}$Faculty of Computing, Harbin Institute of Technology, Harbin, China\\
$^{2}$China Unicom, Beijing, China}
\begin{document}
\ninept
\maketitle

\begin{abstract}
Existing singing pitch correction approaches rely on target melodies or
accompaniment tracks, which may be unavailable in practice. We formulate
reference-free singing pitch correction as a
music-constrained sequence editing task that determines whether and how each
note should be corrected from the input performance alone. A pretrained
symbolic music encoder with lightweight singing-domain adapters produces
contextual representations of the singing MIDI. Based on these representations, 
two lightweight correction heads jointly
model correction necessity and signed pitch modification through a factorized
pitch-editing distribution. An input-dependent tonal prior
derived from the estimated key distribution then reranks the candidate offsets,
favoring tonally compatible corrections without target melodies or
ground-truth key annotations. Experiments on real paired amateur and professional
singing recordings show that the method improves note-level pitch accuracy
from 73.25\% to 83.43\%, outperforming a context-based baseline by 3.99
percentage points while balancing error correction and preservation of
correctly performed notes.

\end{abstract}
\begin{keywords}
Singing pitch correction, reference-free, sequence editing, tonal constraint.
\end{keywords}
\section{Introduction}
\label{sec:intro}

Pitch accuracy is crucial for singing quality, yet amateur performances often exhibit
significant pitch deviations that degrade overall musicality. Automatic
singing pitch correction has become an important component in music production, online singing
platforms, and AI-assisted music creation. However, many practical scenarios, including
user-generated a cappella recordings, spontaneous humming, melody sketching, and legacy
recordings without project files, provide only vocal performance without a 
reference melody. In such cases, reference-free correction must infer intended pitches from
the musical context of the input while preserving expressiveness.

Existing singing pitch correction methods typically rely on explicit musical references or
predefined pitch constraints. Commercial tools such as Auto-Tune and Melodyne provide
scale- or note-constrained pitch editing~\cite{hildebrand_pitch_1999,celemony_melodyne}.
Academic systems likewise exploit auxiliary musical information. For instance, Deep Autotuner predicts
note-wise pitch shifts from the vocal and accompaniment spectrograms~\cite{wager_deep_2020},
KaraTuner conditions pitch prediction on musical notes~\cite{zhuang_karatuner_2022}, and
Neural Singing Voice Beautifier aligns an amateur performance with a professional template
pitch curve~\cite{liu_learning_2022}. More recently, generative models based on conditional
variational autoencoders and diffusion have explored pitch correction under score- or
MIDI-derived conditions~\cite{liu_learning_2022,hai_diff-pitcher_2023,wang_contuner_2024,
li2026srfsvbstyleconsistentsingingvoice}.
These systems assume a target
melody, accompaniment, professional template, or predefined scale.
BERT-APC instead infers pitch corrections from the musical context of
the input performance~\cite{kim_bert-apc_2026}.

Reference-free singing pitch correction is fundamentally different from conventional pitch
estimation. While pitch estimation focuses on extracting sung pitches, reference-free
correction additionally requires determining whether each pitch should be modified. Similar
edit decision formulations have been widely explored in sequence correction tasks, where error
detection and editing operations are explicitly modeled for fine-grained sequence
modification~\cite{omelianchuk_gector_2020,malmi_encode_2019,mallinson_felix_2020}. However, singing pitch
correction introduces unique musical characteristics that differ from general sequence
correction: the validity of a note depends not only on its pitch value but also on melodic
context, tonal structure, and expressive singing characteristics. Music representation models
underscore the importance of contextual modeling for music understanding~\cite{li_mert_2024},
while singing research reveals interactions among pitch, technique, and expressive
style~\cite{zhang_gtsinger_2024}. We therefore model correction as note-level sequence editing informed
by melodic context and tonality.

Based on these observations, we propose a reference-free singing pitch correction framework
that formulates the task as a music-constrained sequence editing problem. Specifically,
we adopt MusicBERT, a pretrained symbolic music representation
model~\cite{zeng_musicbert_2021}, as the backbone encoder and introduce lightweight singing-domain
adapters to obtain contextual representations tailored to singing MIDI. Based on these
representations, we design a hierarchical editing strategy with two lightweight correction heads that 
jointly model correction necessity and pitch modification, explicitly modeling note-level 
editing decisions. Furthermore, we estimate a tonal prior from the input singing performance
and use it to rerank candidate corrections, introducing music-theoretic constraints without
requiring target melodies or ground-truth key annotations.
Audio examples are available at \url{https://dongb001.github.io/RF-SPC/}.

The main contributions of this work are as follows:

\begin{itemize}
\item We propose a reference-free singing pitch correction framework that
models the musical context of the input performance using a frozen pretrained
symbolic music encoder with lightweight singing-domain adapters.

\item We develop a hierarchical correction model with two 
correction heads that jointly model correction necessity and pitch modification.

\item We introduce tonal-aware candidate reranking using a tonal prior
estimated from the input without ground-truth key annotations.
\end{itemize}

\section{METHODOLOGY}
\label{sec:method}
\subsection{Problem formulation}

\begin{figure}[t]
\centering
\includegraphics[width=\linewidth]{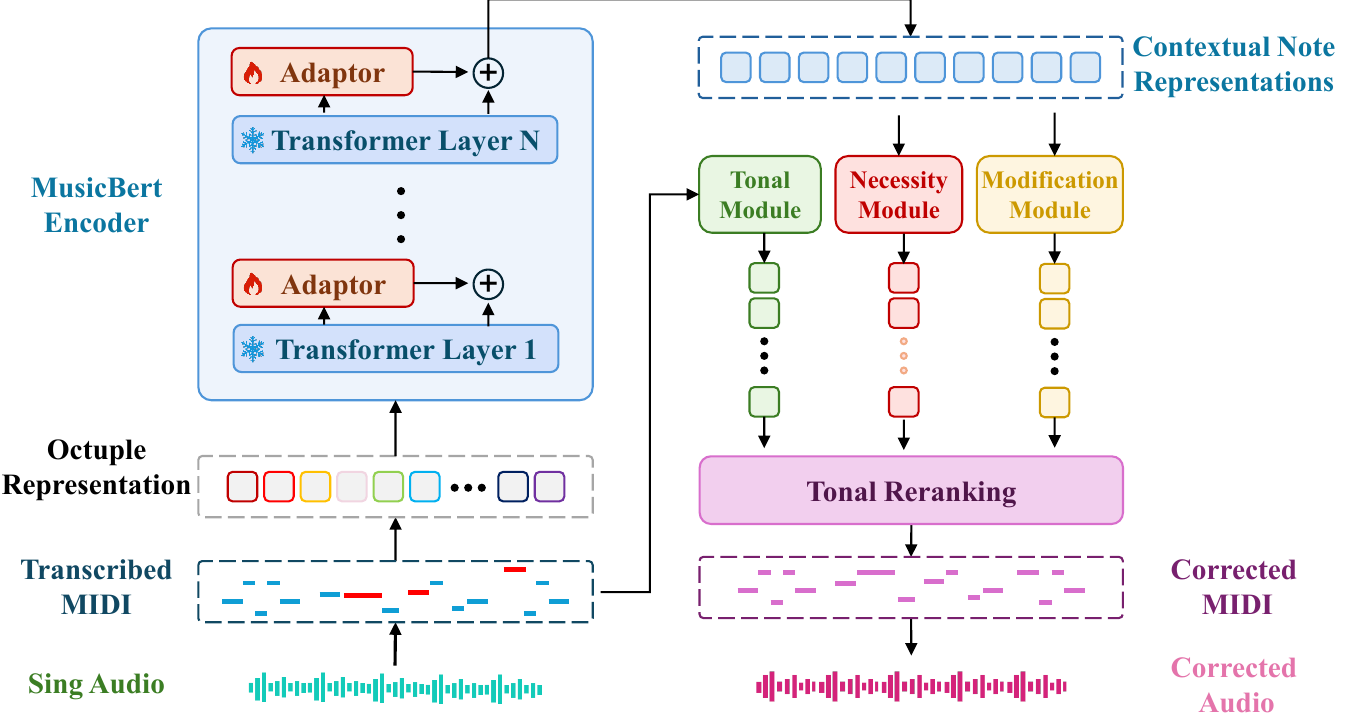}
\caption{Overview of the proposed reference-free singing pitch correction framework.}
\label{fig:architecture}
\end{figure}

We formulate reference-free singing pitch correction in symbolic representation space.
The correction pipeline is illustrated in Fig.~\ref{fig:architecture}. First, SOME~\cite{yu_some_2024} extracts
note-level pitch, onset, and duration from an unaccompanied singing vocal recording,
while RMVPE~\cite{wei_rmvpe_2023} extracts frame-level $F_0$ for reconstruction. To construct
the complete MusicBERT octuple representation, extracted pitch values are quantized
to discrete semitones, the instrument is set to a fixed vocal code, and bar positions,
time signature, and tempo are estimated from the onset sequence. After symbolic
correction, modified pitches are mapped back to $F_0$ contours via semitone-based
pitch shifting on stable regions with linear interpolation at note boundaries, then
resynthesized using NSF-HiFi-GAN~\cite{wang_nsf_2020,kong_hifigan_2020}.

The extracted singing MIDI sequence is represented as
\begin{equation}
\mathcal{X}=\{x_i\}_{i=1}^{N}, \quad
x_i=(p_i,d_i,s_i,e_i),
\label{eq:input-sequence}
\end{equation}
where $N$ denotes the number of detected notes, and $p_i$, $d_i$, $s_i$, and $e_i$ denote
the pitch, duration, onset position, and ending position of the $i$-th note, respectively.
Only pitch is edited while the temporal structure is preserved; for MusicBERT,
$\mathcal{X}$ is augmented with the remaining octuple fields to form
$\mathcal{X}_{\mathrm{oct}}$.

The objective is to generate a corrected singing representation
\begin{equation}
\mathcal{Y}=\{y_i\}_{i=1}^{N}, \quad
y_i=(\hat{p}_i,d_i,s_i,e_i),
\label{eq:output-sequence}
\end{equation}
where $y_i$ is the corrected note and $\hat{p}_i$ is its pitch. The correction process can
therefore be formulated as estimating a pitch modification $\Delta_i=\hat{p}_i-p_i$ applied to the $i$-th note.

Since pitch errors are
typically sparse in singing, an editing-based approach allows selective modification of
erroneous notes while preserving correct ones, avoiding unnecessary alterations that
full sequence prediction would introduce. The correction process consequently requires
an editing decision that indicates whether a note should be modified. We introduce a binary
editing indicator:
\begin{equation}
z_i=
\begin{cases}
1, & \text{if note $i$ requires correction},\\
0, & \text{otherwise}
\end{cases}.
\label{eq:edit-indicator}
\end{equation}
The final corrected pitch is:
\begin{equation}
\hat{p}_i=p_i+\hat{\Delta}_i,
\label{eq:final-pitch}
\end{equation}
where $\hat{\Delta}_i$ denotes the predicted pitch modification. A zero-valued
prediction preserves the input pitch, whereas a nonzero prediction applies a correction.
This formulation frames reference-free singing pitch correction as a music-constrained
sequence editing task, where the model learns when and how to modify pitch deviations
from the intrinsic musical context, guided by hierarchical editing decisions and tonal
priors described in subsequent sections.

\subsection{Singing-aware music context encoder}

Reference-free singing pitch correction requires modeling musical context to infer appropriate
pitch modifications. We therefore employ a pretrained symbolic music encoder to capture
contextual dependencies among musical events. Specifically, we adopt MusicBERT as the backbone
encoder~\cite{zeng_musicbert_2021}, which learns contextual representations of structured
musical events through Transformer-based pretraining on large-scale symbolic MIDI corpora.

Although MusicBERT provides strong symbolic music representations, it is pretrained on general
MIDI data, whereas singing MIDI exhibits different characteristics, including monophonic melodic
structures and vocal-specific pitch variations. To bridge this domain gap, we introduce
lightweight singing-domain adapters~\cite{houlsby_adapter_2019} while keeping the pretrained
MusicBERT parameters frozen. The adapted representation is: 
\begin{equation}
\widetilde{H}_l=H_l+A_l(H_l), \qquad l=1,\ldots,12,
\label{eq:adapted-representation}
\end{equation}
Here, $H_l\in\mathbb{R}^{N\times768}$ denotes the output of the $l$-th
MusicBERT Transformer block, and $h\in\mathbb{R}^{768}$ denotes an
individual token representation. We use
$\widetilde{H}=\widetilde{H}_{12}$ as the final contextual representation.
A gated adapter is inserted after each of MusicBERT's 12 Transformer blocks:
\begin{equation}
A_l(h)=\sigma(g_l)W_{2,l}\operatorname{Drop}_{0.1}\!\left(
\operatorname{GELU}\!\left(W_{1,l}\operatorname{LN}(h)\right)\right),
\label{eq:adapter}
\end{equation}
where $W_{1,l}\in\mathbb{R}^{64\times768}$,
$W_{2,l}\in\mathbb{R}^{768\times64}$, and $g_l$ is a
trainable scalar gate for the adapter after the $l$-th Transformer block. 
The adapted representations are then fed
into the subsequent hierarchical edit prediction module.

\subsection{Hierarchical correction modeling}

Inspired by explicit editing formulations in sequence correction
tasks~\cite{malmi_encode_2019,omelianchuk_gector_2020}, we formulate reference-free singing
pitch correction as a hierarchical correction process. Instead of directly predicting 
the target pitch sequence, the proposed method decomposes 
correction into two stages: predicting correction necessity and predicting pitch modification. 
Given the contextual representation $\widetilde{H}=\{\tilde{h}_i\}_{i=1}^{N}$, we introduce 
two lightweight correction heads, each consisting of a feed-forward projection layer 
followed by an activation function. The correction-necessity head estimates 
the probability that a note requires correction:

\begin{equation}
a_i=P(z_i=1\mid\tilde{h}_i)=\sigma\left(f_{\mathrm{nec}}(\tilde{h}_i)\right).
\label{eq:necessity}
\end{equation}
The pitch-modification head predicts a distribution over the nonzero offset set
$\mathcal{D}_{\ne0}=\{-3,-2,-1,+1,+2,+3\}$, conditioned on correction:
\begin{equation}
b_i(d)=\operatorname{Softmax}\left(f_{\mathrm{mod}}(\tilde{h}_i)\right)_d,
\quad d\in\mathcal{D}_{\ne0}.
\label{eq:offset}
\end{equation}
The two predictions form a factorized pitch-editing distribution:
\begin{equation}
P(\Delta_i=d\mid\tilde{h}_i)=
\begin{cases}
1-a_i, & d=0,\\
a_i b_i(d), & d\neq0
\end{cases}.
\label{eq:hierarchical-distribution}
\end{equation}
The two correction heads are jointly optimized with classification losses:
\begin{equation}
\mathcal{L}=\mathcal{L}_{\mathrm{nec}}+\mathcal{L}_{\mathrm{mod}},
\label{eq:hierarchical-loss}
\end{equation}
where $\mathcal{L}_{\mathrm{nec}}$ and $\mathcal{L}_{\mathrm{mod}}$ are defined as
\begin{equation}
\mathcal{L}_{\mathrm{nec}}=-\frac{1}{|\mathcal{V}|}\sum_{i\in\mathcal{V}}
\left[z_i^*\log a_i+(1-z_i^*)\log(1-a_i)\right],
\label{eq:loss-necessity}
\end{equation}
\begin{equation}
\mathcal{L}_{\mathrm{mod}}=-\frac{1}{|\mathcal{E}|}\sum_{i\in\mathcal{E}}
\log b_i(\delta_i^*).
\label{eq:loss-modification}
\end{equation}
Here $\mathcal{V}$ denotes the set of valid notes, $\mathcal{E}\subseteq\mathcal{V}$
denotes the subset of notes that were perturbed in the training data,
$z_i^*=\mathbf{1}[i\in\mathcal{E}]$ is the binary correction label, and $\delta_i^*$
is the target offset for note $i\in\mathcal{E}$.
The necessity prediction is trained with binary cross-entropy over all notes,
whereas the modification prediction is optimized with cross-entropy only on notes
that require correction.

\subsection{Tonal-aware reranking}

Although the hierarchical correction model predicts pitch modifications from singing context,
reference-free correction can still yield multiple plausible candidates without an explicit
melody reference. We therefore introduce a tonal-aware candidate reranking strategy. Rather
than imposing a hard key constraint, it uses tonal information as a soft prior derived solely
from the input singing performance.

We apply the Krumhansl--Schmuckler (KS) key-finding
algorithm~\cite{krumhansl_cognitive_2001} to the input sequence $\mathcal{X}$. Let
$u_{\mathcal{X}}$ be its duration-weighted 12-bin pitch-class histogram, whose $v$-th bin is
$u_{\mathcal{X}}(v)=\sum_i d_i\mathbf{1}[(p_i\bmod 12)=v]$. Pearson correlations with
the 24 major and minor key templates $\kappa_k$ form a soft key posterior:
\begin{equation}
\rho(k\mid\mathcal{X})=\operatorname{Softmax}_{k}
\left(\operatorname{corr}(u_{\mathcal{X}},\kappa_k)/\tau\right).
\label{eq:key-posterior}
\end{equation}
We denote the resulting posterior vector by
\[
\rho_{\mathcal X}=\{\rho(k\mid\mathcal X)\}_{k=1}^{24}.
\]
We set the posterior temperature to $\tau=0.15$.
The posterior induces an input-dependent pitch-class prior: 
\begin{equation}
r_{\mathcal{X}}(v)=\sum_{k}\rho(k\mid\mathcal{X})\pi_k(v),
\label{eq:pitch-class-prior}
\end{equation}
where $v\in\{0,\ldots,11\}$. For the seven-class scale set $\mathcal{S}_k$,
$\pi_k(v)$ assigns weight 1 to $v\in\mathcal{S}_k$ and $\eta=0.15$ otherwise,
normalized by $7+5\eta$. We compute a song-level confidence $c(\mathcal{X})\in[0,1]$
from the posterior entropy and the input-note count $N$:
\begin{equation}
c(\mathcal{X})=\left(1-\frac{\mathcal{H}(\rho_{\mathcal{X}})}{\log 24}\right)
\min\left(1,\frac{N}{2n_{\min}}\right),
\label{eq:key-confidence}
\end{equation}
where $\mathcal{H}(\rho_{\mathcal{X}})=-\sum_k\rho(k\mid\mathcal{X})
\log\rho(k\mid\mathcal{X})$. We set $n_{\min}=8$; when $N<n_{\min}$ or
$u_{\mathcal{X}}$ has zero variance, we use a uniform posterior and set $c(\mathcal{X})=0$.
For each note, the editor considers the seven offsets
$\mathcal{D}=\{0,\pm1,\pm2,\pm3\}$, where $0$ indicates no modification. 
We combine the editing distribution with the tonal prior using: 
\begin{equation}
\begin{aligned}
S_i(d) &=\log P(\Delta_i=d\mid\tilde{h}_i)\\
&\quad+\lambda c(\mathcal{X})\log r_{\mathcal{X}}((p_i+d)\bmod 12),
\end{aligned}
\label{eq:key-reranking}
\end{equation}
where $\lambda\geq0$ controls the global strength of the tonal constraint and is selected on
the validation set. The final pitch modification is: 
\begin{equation}
\hat{\Delta}_i=\arg\max_{d\in\mathcal{D}}S_i(d).
\label{eq:final-edit}
\end{equation}
The corresponding predicted editing decision is
$\hat{z}_i=\mathbf{1}[\hat{\Delta}_i\neq0]$.
This confidence-weighted reranking biases ambiguous candidates toward tonally compatible
solutions without forcing pitches to conform to a single detected key.

\section{EXPERIMENTAL SETUP}
\label{sec:experimental-setup}

\subsection{Dataset}
\label{sec:dataset}

We construct training samples by extracting singing MIDI
from 16,000 professional recordings (approximately 1,060 hours of Chinese and English popular songs)
and applying synthetic pitch perturbations. Considering the typical vocal range, note pitches are
constrained within MIDI 36--84 (C2--C6). For each sequence, 20\% of notes are perturbed using
the six nonzero offsets in $\mathcal{D}_{\ne0}$. To simulate different error patterns, 
70\% of perturbations are independent, whereas 30\% form correlated spans of two to four notes. 
Synthetic perturbations are used for model development, and real
paired data are reserved exclusively for evaluation.

For evaluation, we use a separate private paired singing dataset containing Chinese and English popular songs.
Each song includes amateur and professional performances recorded by vocal students.
The paired recordings are aligned using dynamic time warping (DTW) to obtain note-level correspondence.
The dataset is split by song, including 20 validation songs (8,299 notes) and 80
held-out test songs (30,682 notes). All hyperparameters, including the tonal reranking
strength, are selected using only the validation set.
The professional performances are used as evaluation references rather than absolute ground truth,
since expressive variations may exist across performances.

\subsection{Implementation Details}
\label{sec:implementation-details}

The pretrained MusicBERT encoder remains frozen throughout
training. We use three training stages: (1) train only the adapters for 8k updates
with MusicBERT's original masked-token prediction head; (2) train only
the two correction heads for 12k updates; and (3) jointly train the
adapters and correction heads for 25k updates. The learning rate is
$10^{-4}$ in Stages 1 and 2; Stage 3 uses $10^{-5}$ for adapters and
$10^{-4}$ for correction heads. We use the AdamW optimizer with cosine annealing schedules.
The correction necessity and pitch modification losses are equally weighted. Each GPU processes
32 sequences with a maximum length of 8,192 field tokens.

\subsection{Baselines}
\label{sec:baselines}

We compare our method with two reference-free
baselines: an independent reimplementation of BERT-APC~\cite{kim_bert-apc_2026} trained on the same corpus, 
and a key-based pitch
snapping method that adjusts notes toward the nearest pitch classes according to the
estimated key. BERT-APC is a recent learning-based approach to this task.

\subsection{Evaluation Metrics}
\label{sec:evaluation-metrics}

We report note-level semitone-accurate pitch accuracy (Acc.) as the primary metric. 
To evaluate correction-preservation trade-offs, we report the error repair rate
($N_{\text{fixed}}/N_{\text{originally wrong}}$) and harm rate
($N_{\text{damaged}}/N_{\text{originally correct}}$), measuring recovery of
incorrect notes and undesired modification of correct notes, respectively.
For subjective evaluation, we assess three perceptual aspects on a five-point Likert scale:
Pitch Correctness (pitch accuracy relative to piano-rendered reference recordings),
Naturalness (vocal quality preservation), and Overall Quality (listening preference).

\section{Results}
\label{sec:results}

We first evaluate overall correction performance, then examine component
contributions, and finally assess perceptual quality.

\subsection{Objective evaluation}
\label{sec:objective-evaluation}

Table~\ref{tab:objective-results} compares the proposed method with two
reference-free baselines on the held-out real test set. The proposed method
improves note-level pitch accuracy from 73.25\% to 83.43\%, increasing agreement with
the professional references on the held-out paired recordings.
It exceeds BERT-APC by 3.99 percentage points, and both gains over the correction
baselines are significant (paired song-level Wilcoxon tests; Holm-adjusted $p<0.001$).
The proposed method also raises the error repair rate from 26.18\% to
46.81\%, while its harm rate remains 3.19\%. 
Key Snapping repairs 20.76\% of the originally incorrect notes but damages
9.62\% of the correct notes, reducing accuracy below the uncorrected input.
A representative pitch-trajectory example is shown in
Fig.~\ref{fig:analysis}(a). 

\begin{table}[!h]
\caption{Objective results on the held-out real test set (\%).}
\label{tab:objective-results}
\centering
\normalsize
\setlength{\tabcolsep}{3pt}
\begin{tabular}{lcccc}
\hline
Method & Acc. $\uparrow$ & $\Delta$Acc. $\uparrow$ & Rep. $\uparrow$ & Harm $\downarrow$\\
\hline
Input & 73.25 & 0.00 & -- & --\\
Key Snapping & 71.76	& -1.49 & 20.76 & 9.62\\
BERT-APC~\cite{kim_bert-apc_2026} & 79.44 & +6.19 & 26.18 & \textbf{1.11}\\
\textbf{Ours} & \textbf{83.43} & \textbf{+10.18} & \textbf{46.81} & 3.19\\
\hline
\end{tabular}
\end{table}

\subsection{Ablation study}
\label{sec:ablation-study}

To isolate the contribution of each component, we evaluate variants without
singing-domain adaptation, hierarchical correction modeling, or tonal-aware
reranking. Table~\ref{tab:ablations} reports the results on the held-out real
test set.

\textbf{Singing-domain adaptation:} We remove the adapter modules and optimize only the
correction heads on top of frozen MusicBERT representations. Removing the adapters reduces
note accuracy from 83.43\% to 76.87\%, showing that adaptation to singing-specific MIDI
characteristics is important for reference-free pitch correction.

\textbf{Hierarchical correction modeling:} The flat edit head variant replaces the 
two correction heads with a single seven-way classifier. 
Compared with the flat head, the hierarchical formulation increases the
repair rate from 39.32\% to 46.81\% and accuracy from 82.35\% to 83.43\%, 
at the cost of a higher harm rate (3.19\% vs.\ 1.93\%).

\begin{figure}[!t]
\centering
\begin{minipage}[b]{0.48\linewidth}
\centering
\includegraphics[width=\linewidth]{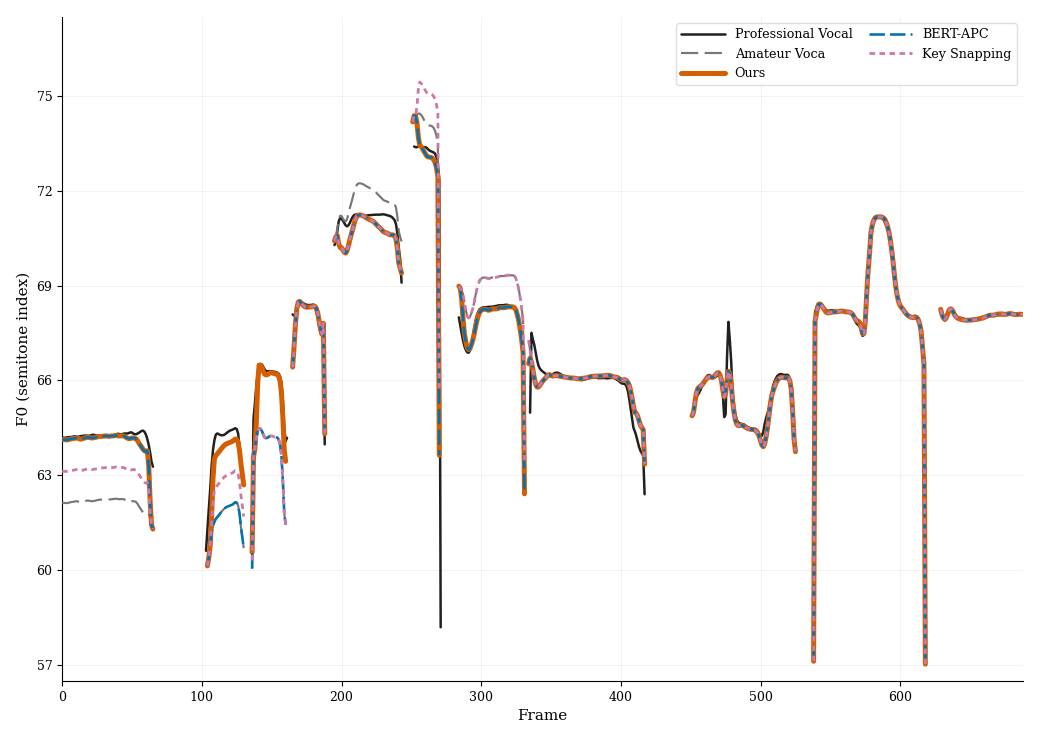}\\[-2pt]
\normalsize (a) Pitch-trajectory example.
\end{minipage}\hfill
\begin{minipage}[b]{0.48\linewidth}
\centering
\includegraphics[width=\linewidth]{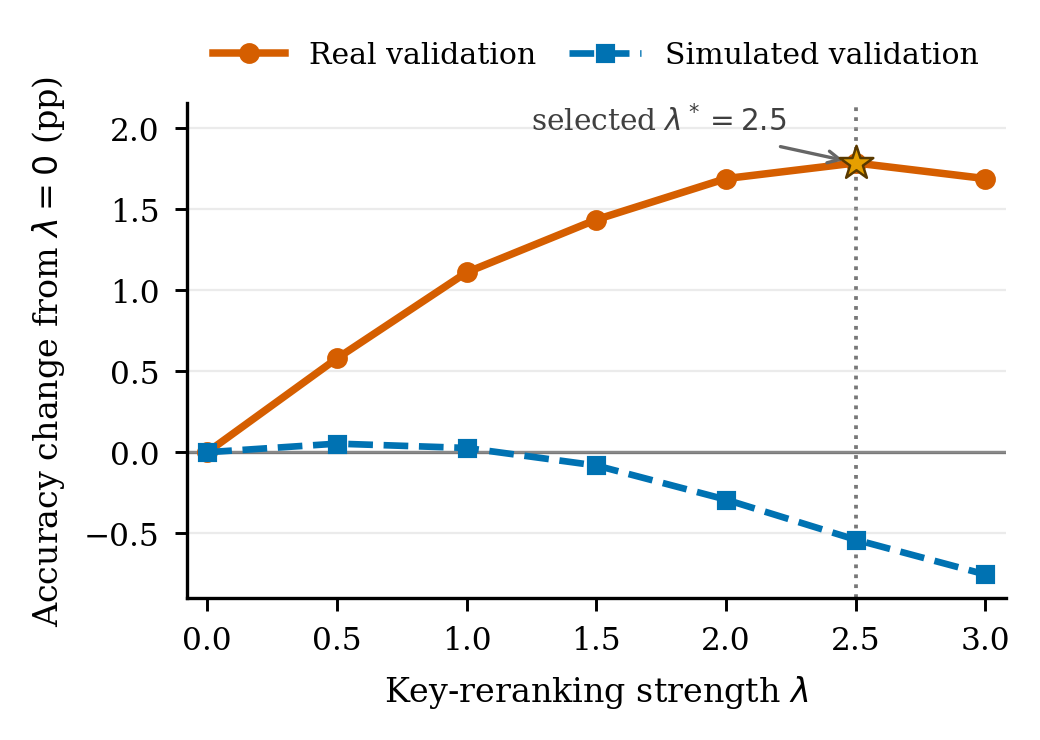}\\[-2pt]
\normalsize (b) Validation response to $\lambda$.
\end{minipage}
\caption{Qualitative and validation analyses.}
\label{fig:analysis}
\end{figure}

\textbf{Tonal-aware reranking:} To evaluate the contribution of tonal information,
we remove the tonal-aware reranking module and compare with the proposed method. 
Removing tonal reranking decreases note-level accuracy from
83.43\% to 81.52\% and the error repair rate from 46.81\% to 35.76\%. This indicates that tonal
information provides useful musical priors for selecting correction candidates.
We further analyze the influence of tonal-constraint strength $\lambda$ on real paired singing 
data, as shown in Fig.~\ref{fig:analysis}(b). Tonal
constraints provide limited improvement on synthetic perturbations, where randomly generated
pitch errors are less correlated with musical structure. In contrast, real singing performances
benefit more from tonal information, indicating that realistic singing deviations have stronger
relationships with the underlying tonal context.
The highest real-validation accuracy is obtained at $\lambda=2.5$.
Accuracy decreases when the tonal term is weighted more strongly, suggesting
that an overly large weight may overemphasize the estimated tonal prior. 
This observation motivates the use of a moderately weighted soft tonal prior.

\begin{table}[!t]
\caption{Ablation results on the held-out real test set (\%).}
\label{tab:ablations}
\centering
\normalsize
\setlength{\tabcolsep}{3pt}
\begin{tabular}{lcccc}
\hline
Method & Acc. $\uparrow$ & $\Delta$Acc. $\uparrow$ & Rep. $\uparrow$ & Harm $\downarrow$\\
\hline
w/o adapters & 76.87 & +3.62 & 34.10 & 7.52\\
Flat edit head & 82.35 & +9.10 & 39.32 & 1.93\\
w/o reranking & 81.52 & +8.27 & 35.76 & \textbf{1.77}\\
\textbf{Ours} & \textbf{83.43} & \textbf{+10.18} & \textbf{46.81} & 3.19\\
\hline
\end{tabular}
\end{table}

\subsection{Subjective evaluation}
\label{sec:subjective-evaluation}

To evaluate the perceptual quality of the corrected singing audio, we conduct a subjective listening
test with 30 vocally trained participants. Each participant evaluates 5 randomly selected singing
excerpts (8--12 seconds) across four methods. Participants are provided with piano-rendered
professional recordings as pitch references before rating. All methods use identical SOME transcription and
NSF-HiFi-GAN reconstruction; only symbolic correction differs.

As shown in Table~\ref{tab:subjective-results}, the proposed method achieves the highest mean scores 
for Pitch Correctness and Overall Quality, improving them from 3.10 to 3.59 and from 3.25 to 3.94 over 
the input, respectively. Its Naturalness score remains close to the input (3.93 vs. 3.97), 
indicating that the perceptual improvements are achieved without substantially degrading naturalness.

\begin{table}[!ht]
\caption{Subjective evaluation on a five-point scale.}
\label{tab:subjective-results}
\centering
\normalsize
\setlength{\tabcolsep}{3pt}
\begin{tabular}{lccc}
\hline
Method & Pitch Corr. $\uparrow$ & Nat. $\uparrow$ & Overall $\uparrow$\\
\hline
Input & 3.10 & \textbf{3.97} & 3.25\\
Key Snapping & 3.26 & 3.80 & 3.67\\
BERT-APC & 3.46 & 3.85 & 3.47\\
\textbf{Ours} & \textbf{3.59} & 3.93 & \textbf{3.94}\\
\hline
\end{tabular}
\end{table}

\section{CONCLUSIONS}
\label{sec:conclusions}

We formulate reference-free singing pitch correction as a music-constrained 
sequence editing task by combining a pretrained symbolic music encoder with 
lightweight singing-domain adapters, hierarchical correction modeling, and tonal-aware reranking. 
Evaluations on real paired amateur and professional recordings show improved
pitch correction while largely preserving naturalness. 
Due to the inherent diversity of musical expression, reference-free correction 
produces one musically plausible interpretation among multiple valid alternatives. 
Future work will extend the framework to jointly model pitch, rhythm, and singing 
techniques for more comprehensive singing voice refinement.

\bibliographystyle{IEEEbib}
\bibliography{Template}

@misc{hildebrand_pitch_1999,
  title={Pitch Detection and Intonation Correction Apparatus and Method},
  author={Hildebrand, Harold A.},
  howpublished={U.S. Patent 5,973,252},
  month={October},
  year={1999},
  url={https://patents.google.com/patent/US5973252A/en}
}

@misc{celemony_melodyne,
  title={What Is {Melodyne}? Note-Based Audio Editing},
  author={{Celemony Software GmbH}},
  howpublished={Online technical documentation},
  note={Accessed: Aug. 22, 2026},
  url={https://www.celemony.com/en/melodyne/what-is-melodyne}
}

@inproceedings{wang_contuner_2024,
  title={{ConTuner}: Singing Voice Beautifying with Pitch and Expressiveness Condition},
  author={Wang, Jianzong and Li, Pengcheng and Zhang, Xulong and Cheng, Ning and Xiao, Jing},
  booktitle={2024 International Joint Conference on Neural Networks (IJCNN)},
  pages={1--6},
  year={2024},
  doi={10.1109/IJCNN60899.2024.10650210}
}

@inproceedings{wager_deep_2020,
  title={Deep {Autotuner}: A Pitch Correcting Network for Singing Performances},
  author={Wager, Sanna and Tzanetakis, George and Wang, Cheng-i and Kim, Minje},
  booktitle={ICASSP 2020--2020 IEEE International Conference on Acoustics, Speech and Signal Processing (ICASSP)},
  pages={246--250},
  year={2020},
  doi={10.1109/ICASSP40776.2020.9054308}
}

@misc{zhuang_karatuner_2022,
  title={{KaraTuner}: Towards End-to-End Natural Pitch Correction for Singing Voice in Karaoke},
  author={Zhuang, Xiaobin and Yu, Huiran and Zhao, Weifeng and Jiang, Tao and Hu, Peng},
  year={2022},
  eprint={2110.09121},
  archivePrefix={arXiv},
  primaryClass={cs.SD},
  doi={10.48550/arXiv.2110.09121}
}

@inproceedings{hai_diff-pitcher_2023,
  title={{Diff-Pitcher}: Diffusion-Based Singing Voice Pitch Correction},
  author={Hai, Jiarui and Elhilali, Mounya},
  booktitle={2023 IEEE Workshop on Applications of Signal Processing to Audio and Acoustics (WASPAA)},
  pages={1--5},
  year={2023},
  doi={10.1109/WASPAA58266.2023.10248127}
}

@article{kim_bert-apc_2026,
  title={{BERT-APC}: A Reference-Free Framework for Automatic Pitch Correction via Musical Context Inference},
  author={Kim, Sungjae and Na, Kihyun and Choi, Jinyoung and Kim, Injung},
  journal={IEEE Transactions on Audio, Speech and Language Processing},
  volume={34},
  pages={2649--2662},
  year={2026},
  doi={10.1109/TASLPRO.2026.3688391}
}

@inproceedings{liu_learning_2022,
  title={Learning the Beauty in Songs: Neural Singing Voice Beautifier},
  author={Liu, Jinglin and Li, Chengxi and Ren, Yi and Zhu, Zhiying and Zhao, Zhou},
  booktitle={Proceedings of the 60th Annual Meeting of the Association for Computational Linguistics (Volume 1: Long Papers)},
  pages={7970--7983},
  year={2022},
  publisher={Association for Computational Linguistics},
  doi={10.18653/v1/2022.acl-long.549}
}

@inproceedings{malmi_encode_2019,
  title={Encode, Tag, Realize: High-Precision Text Editing},
  author={Malmi, Eric and Krause, Sebastian and Rothe, Sascha and Mirylenka, Daniil and Severyn, Aliaksei},
  booktitle={Proceedings of the 2019 Conference on Empirical Methods in Natural Language Processing and the 9th International Joint Conference on Natural Language Processing (EMNLP-IJCNLP)},
  pages={5054--5065},
  year={2019},
  publisher={Association for Computational Linguistics},
  doi={10.18653/v1/D19-1510}
}

@inproceedings{omelianchuk_gector_2020,
  title={{GECToR}---Grammatical Error Correction: Tag, Not Rewrite},
  author={Omelianchuk, Kostiantyn and Atrasevych, Vitaliy and Chernodub, Artem and Skurzhanskyi, Oleksandr},
  booktitle={Proceedings of the Fifteenth Workshop on Innovative Use of NLP for Building Educational Applications},
  pages={163--170},
  year={2020},
  publisher={Association for Computational Linguistics},
  doi={10.18653/v1/2020.bea-1.16}
}

@inproceedings{mallinson_felix_2020,
  title={{FELIX}: Flexible Text Editing Through Tagging and Insertion},
  author={Mallinson, Jonathan and Severyn, Aliaksei and Malmi, Eric and Garrido, Guillermo},
  booktitle={Findings of the Association for Computational Linguistics: EMNLP 2020},
  pages={1244--1255},
  year={2020},
  publisher={Association for Computational Linguistics},
  doi={10.18653/v1/2020.findings-emnlp.111}
}

@inproceedings{li_mert_2024,
  title={{MERT}: Acoustic Music Understanding Model with Large-Scale Self-supervised Training},
  author={Li, Yizhi and Yuan, Ruibin and Zhang, Ge and Ma, Yinghao and Chen, Xingran and Yin, Hanzhi and Xiao, Chenghao and Lin, Chenghua and Ragni, Anton and Benetos, Emmanouil and Gyenge, Norbert and Dannenberg, Roger and Liu, Ruibo and Chen, Wenhu and Xia, Gus and Shi, Yemin and Huang, Wenhao and Wang, Zili and Guo, Yike and Fu, Jie},
  booktitle={International Conference on Learning Representations},
  pages={12181--12204},
  year={2024},
  url={https://proceedings.iclr.cc/paper_files/paper/2024/file/33dffa2e3d2ab74a783d1a8c292f66d9-Paper-Conference.pdf}
}

@inproceedings{zhang_gtsinger_2024,
  title={{GTSinger}: A Global Multi-Technique Singing Corpus with Realistic Music Scores for All Singing Tasks},
  author={Zhang, Yu and Pan, Changhao and Guo, Wenxiang and Li, Ruiqi and Zhu, Zhiyuan and Wang, Jialei and Xu, Wenhao and Lu, Jingyu and Hong, Zhiqing and Wang, Chuxin and Zhang, Lichao and He, Jinzheng and Jiang, Ziyue and Chen, Yuxin and Yang, Chen and Zhou, Jiecheng and Cheng, Xinyu and Zhao, Zhou},
  booktitle={Advances in Neural Information Processing Systems 37},
  pages={1117--1140},
  year={2024},
  doi={10.52202/079017-0034}
}

@inproceedings{zeng_musicbert_2021,
  title={{MusicBERT}: Symbolic Music Understanding with Large-Scale Pre-Training},
  author={Zeng, Mingliang and Tan, Xu and Wang, Rui and Ju, Zeqian and Qin, Tao and Liu, Tie-Yan},
  booktitle={Findings of the Association for Computational Linguistics: ACL-IJCNLP 2021},
  pages={791--800},
  year={2021},
  publisher={Association for Computational Linguistics},
  doi={10.18653/v1/2021.findings-acl.70}
}

@inproceedings{houlsby_adapter_2019,
  title={Parameter-Efficient Transfer Learning for {NLP}},
  author={Houlsby, Neil and Giurgiu, Andrei and Jastrzebski, Stanislaw and Morrone, Bruna and De Laroussilhe, Quentin and Gesmundo, Andrea and Attariyan, Mona and Gelly, Sylvain},
  booktitle={Proceedings of the 36th International Conference on Machine Learning},
  pages={2790--2799},
  year={2019},
  volume={97},
  series={Proceedings of Machine Learning Research},
  publisher={PMLR},
  url={https://proceedings.mlr.press/v97/houlsby19a.html}
}

@book{krumhansl_cognitive_2001,
  title={Cognitive Foundations of Musical Pitch},
  author={Krumhansl, Carol L.},
  year={2001},
  month=nov,
  publisher={Oxford University Press},
  doi={10.1093/acprof:oso/9780195148367.001.0001},
  url={https://doi.org/10.1093/acprof:oso/9780195148367.001.0001}
}

@misc{yu_some_2024,
  title = {{SOME}: Singing-Oriented {MIDI} Extractor},
  author = {{OpenVPI}},
  howpublished = {GitHub repository},
  note = {Accessed: Sep. 24, 2026},
  url = {https://github.com/openvpi/SOME}
}

@inproceedings{wei_rmvpe_2023,
  title = {{RMVPE}: A Robust Model for Vocal Pitch Estimation in Polyphonic Music},
  author = {Wei, Haojie and Cao, Xueke and Dan, Tangpeng and Chen, Yueguo},
  booktitle = {Interspeech 2023},
  pages = {5421--5425},
  year = {2023},
  doi = {10.21437/Interspeech.2023-528}
}

@article{wang_nsf_2020,
  title = {Neural Source-Filter Waveform Models for Statistical Parametric Speech Synthesis},
  author = {Wang, Xin and Takaki, Shinji and Yamagishi, Junichi},
  journal = {IEEE/ACM Transactions on Audio, Speech, and Language Processing},
  volume = {28},
  pages = {402--415},
  year = {2020},
  doi = {10.1109/TASLP.2019.2956145}
}

@inproceedings{kong_hifigan_2020,
  title = {{HiFi-GAN}: Generative Adversarial Networks for Efficient and High Fidelity Speech Synthesis},
  author = {Kong, Jungil and Kim, Jaehyeon and Bae, Jaekyoung},
  booktitle = {Advances in Neural Information Processing Systems},
  volume = {33},
  pages = {17022--17033},
  year = {2020}
}

@misc{li2026srfsvbstyleconsistentsingingvoice,
  title={{{SRF-SVB}}: Style-Consistent Singing Voice Beautifying via Rectified Flow},
  author={Li, Wenhui and Dong, Biao and Hu, Liwei and Han, Jiqing and He, Yongjun},
  year={2026},
  eprint={2609.25610},
  archivePrefix={arXiv},
  primaryClass={eess.AS},
  url={https://arxiv.org/abs/2609.25610}
}

\end{document}